\documentclass[prc,twocolumn]{revtex4-2}
\pdfoutput=1
\usepackage{graphicx}
\usepackage{lineno}
\usepackage{amssymb}
\usepackage{amsmath}
\usepackage{bm}
\usepackage{adjustbox}

\begin{document}

\title[]{Investigating fission dynamics of neutron shell closed nuclei $^{210}$Po, $^{212}$Rn and $^{213}$Fr within a stochastic dynamical approach}

\author{Divya Arora}
\author{ P. Sugathan}
\email{sugathan@gmail.com}
\author{A. Chatterjee}
\affiliation{Inter-University Accelerator Centre, Aruna Asaf Ali Marg, New Delhi 110067, India }

\begin{abstract}
Dissipative dynamics of nuclear fission is a well confirmed phenomenon described either by a Kramers-modified statistical model or by a dynamical model employing the Langevin equation. Though dynamical models as well as statistical models incorporating fission delay are found to explain the measured fission observables in many studies, it nonetheless shows conflicting results for shell closed nuclei in the mass region 200. Analysis of recent data for neutron shell closed nuclei in excitation energy range 40$-$80 MeV failed to arrive at a satisfactory description of the data and attributed the mismatch to shell effects and/or entrance channel effects, without reaching a definite conclusion.  In the present work we show that a well established stochastic dynamical code simultaneously reproduces the available data of pre-scission neutron multiplicities, fission and evaporation residue excitation functions for neutron shell closed nuclei $^{210}$Po and $^{212}$Rn  and their isotopes $^{206}$Po and $^{214,216}$Rn without the need for including any extra shell or entrance channel effects. The calculations are performed by using a phenomenological universal friction form factor with no ad-hoc adjustment of model parameters. However, we note significant deviation, beyond experimental errors, in some cases of Fr isotopes.

\end{abstract}

%

%
\maketitle

\section{Introduction}
Fission of atomic nuclei is considered to be one of the most complex physical phenomena in nuclear physics. It involves rapid re-arrangement of nuclear matter with a delicate interplay between the macroscopic bulk matter and the microscopic quantal properties \cite{Vandenbosch1973,Krappe2012}. Though properties of fission have been studied exhaustively, many aspects of the dynamics are still not well-understood. For instance, discrepancies are reported between the measured fission observables and the predictions of the classical theory based on the standard Bohr-Wheeler  statistical model of fission \cite{Bohr1939}. Fission hindrance, enhanced pre-scission particle and giant dipole resonance (GDR) $\gamma$-ray multiplicities observed in hot nuclei suggested the effects of nuclear dissipation slowing down the fission process \cite{Gavron1981,Newton1988,Hinde1986,Lestone1991,Rossner1989,Thoennessen1987,Hofman1995}. To account for frictional effects, Kramers diffusion model formalism with modified fission width \cite{Kramers1940}, referred to as Kramers-modified statistical model was included in the standard statistical theory.

Although nature and strength of the nuclear dissipation have been studied quite extensively, a simultaneous description of the experimental observables, namely, pre-scission neutron multiplicities ($\nu_{pre}$), fission excitation functions and evaporation residue (ER) cross-sections still remains challenging. Additionally, the  dissipation coefficient is treated as an adjustable free parameter  in the statistical model analysis. The pre-fission lifetime (or the dissipation strength), the level density parameter at ground state and saddle point deformation and fission barrier are empirically fitted to explain the $\nu_{pre}$ and/or fission and ER cross-section data  \cite{Hinde1986,Mahata2006,Mancusi2010,Saxena1994,Singh2014}. As a result, the conclusions reached are often system dependent and are inadequate to provide a consistent description of the fission process. 

Inadequate modelling of fission in statistical model can drastically influence the understanding of the fission phenomenon \cite{McCalla2008,Lestone2009}. This is especially observed in mass (A) $\approx$ 200 region that is explored here, to understand the role of N$=$126 neutron shell closure in the fissioning compound nucleus (CN). An anomalous increase in the experimental fission fragment angular anisotropy was reported for $^{210}$Po (N$=$126) as compared to $^{206}$Po (non-shell closed nuclei) across an  excitation energy range ($E_{ex}) \approx$ 40$-$60 MeV and was conjectured to be a manifestation of shell effects at the unconditional saddle \cite{Shrivastava1999}. Further, a considerable amount of saddle shell correction was invoked to describe the experimental $\nu_{pre}$ data for $^{210}$Po nuclei \cite{Golda2013}. However, a re-investigation of the experimental excitation functions and $\nu_{pre}$ data of $^{210}$Po ruled out any significant shell influence on the saddle \cite{Mahata2015} after correlated tuning of statistical-model parameters and inclusion of fission delay.

Another interesting aspect  is the contradictory interpretation for correlation between neutron shell structure  and nuclear dissipation strength that was required to reproduce the measured ER  and $\nu_{pre}$ excitation functions in N$=$126 shell closed nuclei, namely $^{212}$Rn and  $^{213}$Fr.  The theoretical analysis of $\nu_{pre}$ data of $^{212}$Rn \cite{Sandal2013} and $^{213}$Fr \cite{Singh2012} reported a low dissipation strength at $E_{ex}\approx$ 50 MeV which was attributed to the influence of neutron shell closure. On the contrary, no discernible shell influence was reported from ER cross-section studies of $^{212}$Rn and its isotope \cite{Laveen2015}, though moderate nuclear dissipation was required to describe the data. It must be noted that the magnitude of  dissipation invoked to explain the experimental ER cross-sections varied within Rn isotopes \cite{Laveen2015,Sandal2015}, which is again found to be different for the description of the $\nu_{pre}$ data \cite{Sandal2013}. Interestingly, in case of Fr nuclei, the finite-range liquid drop model fission barrier was scaled down, particularly for $^{213}$Fr to fit  measured ER cross-section \cite{Singh2014}. This reduction of the fission barrier is in disagreement with the predictions for the shell closed nuclei \cite{Moeller2009}. One notable observation is the reported interpretation of reduced survival probability of  $^{213}$Fr nucleus due to neutron shell which is in contrast to the isotopic trend reported for Rn isotopes. Further, the fission cross-section of $^{213}$Fr is reported to exhibit no extra stability from N=126 shell closure \cite{Singh2021}. In the statistical model approach followed in these works, no attempts were made to extract a global prescription of the parameters, rather, a case specific adjustment of dissipation strength was involved. The influence of neutron shell structures on the potential energy surface and hence fission observables are still quite ambiguous. Apart from just shell influence, entrance channels effects are also probed in a couple of recent publications to understand the experimental $\nu_{pre}$ data for $^{213}$Fr nuclei \cite{Shareef2016,Schmitt2018}. These studies reportedly observed a deviation in the measured data from the predictions of entrance channel model for $^{16}$O- and $^{19}$F-induced reactions.

These studies substantiate the view that no consistent picture has emerged from recent independent analysis  of each fission observable for neutron shell closed nuclei $^{210}$Po, $^{212}$Rn and $^{213}$Fr and their isotopes. Inadequacies of standard statistical model interpretations have been addressed by employing Kramers-modified fission width taking into account shape-dependent level density, temperature-dependent fission transition points, orientation (K state) degree of freedom and temperature-independent reduced dissipation coefficient \cite{McCalla2008,Lestone2009}. Attempts for restraining the statistical-model parameters have also been reported \cite{Banerjee2018}, but a consistent description of experimental data for all three observables, namely $\nu_{pre}$, fission and ER excitation functions for shell closed nuclei still could not be achieved. Recent developments in  multi-dimensional stochastic approach are fairly successful in describing the fission characteristics of excited nuclei \cite{Karpov,Nadtochy2002,Randrup2011,Randrup2011a,Nadtochy2014,Mazurek2017}. However, a simultaneous description of the experimental data and a systematic study for shell closed nuclei has not been attempted yet and is further required.

In this paper, we show that the dynamical model based on 1D Langevin equation coupled with a statistical approach \cite{Frobrich1998} can simultaneously reproduce $\nu_{pre}$, fission and ER cross-section data  of shell closed nuclei over a range of excitation energies ($E_{ex} \approx$ 40$-$80 MeV) of the measurements. The present calculations are performed without adjusting any of the model parameters, thus provides a unified framework for a simultaneous study of these fission observables for nuclei in A $\approx$ 200 region. We re-investigated the available experimental data for the neutron shell closed nuclei  $^{210}$Po, $^{212}$Rn and $^{213}$Fr, and their non-shell closed isotopes $^{206}$Po, $^{214,216}$Rn and $^{215,217}$Fr.  It is observed that a universal deformation-dependent reduced friction parameter is able to describe the fission observables simultaneously at all measured energies irrespective of the shell structure of the nuclei.

\section{Theoretical Model Description}
A combined dynamical and  statistical model code \cite{code} is utilized to compute the fission observables of nuclei under study. The detailed description of the theoretical aspects of the model can be found in Refs. \cite{Frobrich1998,Gontchar1993}. The dynamical part of the model is carried out with a 1D Langevin equation of motion governed by a driving potential that is determined by free energy F(q, T), as employed in recent Refs. \cite{Chaudhuri2003,Kun2005,Wei2008,Feng2010,Wang2018,Wang2021}. The free energy as derived from the Fermi gas model is related to the deformation dependent level density parameter a(q, A) as F(q, T) = V(q) - a(q, A)$T^2$ where $T$ is the nuclear temperature, $q$ is the dimensionless deformation coordinate defined as the ratio of half the distance between the center of masses of future fission fragments to the radius of CN and $V(q)$ is the nuclear potential energy obtained from the finite-range liquid drop model \cite{Myers1966,Krappe}. Fr{\"{o}}brich \cite{Frobrich2007} and Lestone $et$ $al$. \cite{Lestone2009} have emphasized on using nuclear entropy given by, S(q, A, $E_{tot}$) = 2 $\sqrt{a(q, A)[E_{tot}-V(q)]}$ in determining the driving force and therefore, it is employed as a crucial quantity in the model. The nuclear driving force K = -$\frac{dV(q)}{dq}$ + $\frac{da(q)}{dq}T^{2}$, not only consists of a conservative force but also contain a thermodynamical correction that enters the dynamics via. level density parameter a(q, A). The deformation dependent level density parameter used in constructing the entropy has the form \cite{Balian1970}:
\begin{equation}
a(q,A) = \tilde{a_1}A + \tilde{a_2}A^{2/3}B_s(q)
\end{equation}
 where $A$ is the mass number of the CN and  $\tilde{a_1}$ = 0.073 MeV$^{-1}$ and  $\tilde{a_2}$ = 0.095  MeV$^{-1}$ are taken from Ref. \cite{Ignatyuk1975}. B$_s$(q) is the dimensionless functional of the surface energy \cite{Gontchar1993,Chaudhuri2002,Nadtochy2014,Wang2018}, expressed as the ratio of surface energy of the composite system to that of a sphere.

The over-damped Langevin equation which describes the fission process in the dynamical part of the model thus, has the form \cite{Frobrich1998}:
\begin{equation}
\frac{dq}{dt} = \frac{T}{M\beta(q)} \left[\frac{\partial S(q)}{\partial q} \right]_ {{E}_{tot}}+ \sqrt{\frac{T}{M\beta(q)}} \Gamma(t)
\end{equation}
where  ${E}_{tot}$ is the total energy of the composite system that remains conserved and $\Gamma(t)$ is a Markovian stochastic variable with a normal distribution. The reduced dissipation coefficient $\beta(q)$ = $\gamma$/M (as employed in literature, see e.g., Refs. \cite{McCalla2008,Banerjee2018,Feng2010,Wang2021} (and Refs. therein)) is the ratio of friction coefficient $\gamma$ to the inertia parameter M calculated with Werner-Wheeler approximation of an incompressible irrotational fluid \cite{Davies}. The present model employs "funny$-$hills" parameters \{c,h,$\alpha$\} \cite{BRACK} for describing the shape of the fissioning nuclei. Taking into account only symmetric fission, the mass asymmetry parameter of the shape evolution is set to $\alpha$=0 \cite{Gontchar1993,Frobrich1998,Chaudhuri2002}. The dimensionless fission coordinate (q) is given by q(c,h)= ($\frac{3c}{8}$)(1+$\frac{2}{15}$[2h+$\frac{(c-1)}{2}$]c$^{3}$), where c and h defines the elongation and neck degree of freedom of the fissioning nucleus, respectively \cite{Frobrich1998,Wang2018,Ye2016,Ye2015}.

Following the fission dynamics through full Langevin dynamical calculation is quite time consuming. Similar to previous Langevin studies \cite{Frobrich1998,Nadtochy2002,Chaudhuri2003,Kun2005,Wei2008,Feng2010,Wang2018}, a computationally less intensive approach is adopted in present study where the dynamical stage is coupled with a statistical model.  In the present calculations, the emission of light particles from ground state to scission configuration along the Langevin trajectories is treated as a discrete process. The evaporation of pre-scission light particles from ground state of Langevin trajectories to the scission point is coupled to the fission mode by a Monte Carlo procedure. The decay width for light particle evaporation at each Langevin time step is calculated with the formalism as suggested by Fr{\"{o}}brich $et$ $al$. \cite{Frobrich1998} and later incorporated in Refs. \cite{Nadtochy2014,Kun2005,Wei2008,Feng2010,Wang2018}. The emission width of a particle of kind $\nu$ (n,p,$\alpha$) is given by \cite{Blann1980}:
\begin{multline}
\Gamma_{\nu} = (2s_{\nu}+1) \frac{m_{\nu}}{\pi^{2}\hbar^{2}\rho_{c}(E_{ex})} \\
 \times \int_{0}^{(E_{ex}-B_{\nu})} d\epsilon_{\nu}\rho_{R}(E_{ex}-B_{\nu}-\epsilon_{\nu})\epsilon_{\nu}\sigma_{inv}(\epsilon_{\nu})
\end{multline}
where s$_{\nu}$ is the spin of emitted particle $\nu$, and m$_{\nu}$ is its reduced mass with respect to the residual nucleus. The level densities of the compound and residual nuclei are denoted by $\rho_{c}$(E$_{ex}$) and $\rho_{R}$(E$_{ex}-B_{\nu}-\epsilon_{\nu}$). B$_{\nu}$ is the liquid-drop binding energy, $\epsilon$ is the kinetic energy of the emitted particle and $\sigma_{inv}(\epsilon_{\nu}$) is
the inverse cross sections \cite{Blann1980}. The decay width for light particle emission is calculated at each Langevin time step $\tau$ \cite{Wang2018,Ye2016,Ye2015}. 

When a stationary flux over the barrier is reached after a sufficiently long delay time, the decay of the CN is then modelled by an adequately modified statistical model \cite{Gontchar1993,Mavlitov1992,Frobrich1993}. To have continuity when switching from dynamical to statistical branch, an entropy dependent fission width is incorporated in the latter. While entering the statistical branch, the particle emission width $\Gamma_{\nu}$ is re-calculated and the fission width  $\Gamma_{f}$ = $\hbar$R$_{f}$ \cite{Frobrich1998} is calculated with fission rate (R$_{f}$) given by, 
\begin{multline}
R_{f} = \frac{T_{gs}\sqrt{|S^{''}_{gs}|S^{''}_{sd}}}{2\pi M \beta_{gs}}exp[S(q_{gs})-S(q_{sd})] \\
\times 2(1+ $erf$[(q{sc}-q{sd})\sqrt{S^{''}_{sd}/2}])^{-1}
\end{multline}
Here erf(x) = (2/$\sqrt{\pi}$) $\int_{0}^{x}$ dt exp($-t^{2}$) is the error function and $\beta_{gs}$ is ground state dissipation coefficient. The saddle-point (q$_{sd}$) and the ground-state positions (q$_{gs}$) are defined by the entropy and not, as in the conventional approach, by the potential energy. The standard Monte Carlo cascade procedure was used to select the kind of decay  with  weights $\Gamma_{i}$/$\Gamma_{tot}$ ($i$=fission,n,p,d,$\alpha$) and $\Gamma_{tot}$  = $\sum_{i}\Gamma_{i}$.  Pre-scission particle multiplicities are calculated by counting the number of corresponding evaporated particle events registered in the dynamical and statistical branch of the model.

The Langevin equation is started from a ground state configuration with a temperature corresponding to the initial excitation energy.  The fusion cross-section can be determined from the partial cross section $\frac{d\sigma(l)}{dl}$  which represent the contribution of angular
momenta $\l$ to the total fusion cross-section. Each Langevin trajectory is started with an orbital angular momentum which is sampled from a fusion spin distribution that reads as \cite{Frobrich1998,Nadtochy2014}:
\begin{equation}
\frac{d\sigma(l)}{dl} = \frac{2\pi}{k^{2}}\frac{2l+1}{1+\exp{\frac{(l-l_{c})}{\delta l}}}
\end{equation}
The final results are weighted over all relevant waves, that is, the spin distribution is used as an angular momentum weight function with which the Langevin calculations for fission are started. As shown in  recent Langevin studies, \cite{Chaudhuri2003,Kun2005,Wei2008,Feng2010,Nadtochy2014,Wang2018,Wang2021}, the spin distribution is calculated with the surface friction model \cite{Marten1992}. This calculation also fixes the fusion cross-section thus guaranteeing the correct normalization of fission and evaporation residue cross-sections within the accuracy of the surface friction model. The parameters $l_{c}$ and $\delta l$ are the critical angular momentum for fusion and diffuseness, respectively.

The fission observables that will be discussed in subsequent sections are calculated in the model as follows. The pre-scission neutron multiplicity is the number of neutrons emitted by the CN till it reaches the scission configuration. The fission probability ($P_{f}$) is given by the ratio of fissioned trajectories to total  trajectories. The CN survival probability (1-$P_{f}$) is given by number of trajectories leading to ER formation divided by total trajectories and the fission (ER) cross-section  is given by the product of  fission (survival) probability and fusion cross-section.

\section{Results and Discussion}

 In the present study, pre-scission neutron multiplicities, fission and ER excitation functions for $^{206,210}$Po, $^{212,214,216}$Rn and $^{213,215,217}$Fr compound nuclei are computed and compared with available experimental data  wherein $^{210}$Po, $^{212}$Rn and $^{213}$Fr are N=126 neutron shell closed nuclei. The table \ref{table:table1} shows important parameters for the reactions studied in this work.  The dynamical calculations are performed with a universal frictional form of Refs. \cite{Frobrich1993,Frobrich1998,Frobrich2007} without adjusting any of the model parameters with a consistent prescription of the dissipation coefficient.  To account for sufficient statistics, $10^7$ Langevin trajectories are considered in the model calculations.


\begin{table*}[t]
\caption{Important parameters of reactions studied}
\label{table:table1}
\centering
\begin{adjustbox}{width=0.7\textwidth}
\small
\begin{tabular}{c c c c c c c c c}
 
\hline
 \hline\hline
\multicolumn{1}{c} {CN} & fissility & {S$_{n}$} &{B$_{f}$($\l$=0)} &{Reaction} &   \multicolumn{2}{c}{Mass excess (MeV)}  &  $\alpha$/$\alpha_{BG}$  \\

\cline{6-7}
      & & (MeV) & (MeV)   &         & target(proj)  & CN   &\\  [0.5ex]  
 \hline
 $^{206}$Po & 0.717   & 7.99 & 10.51  & $^{12}$C+$^{194}$Pt& -34.79(0)   &  -18.83 &  1.043\\
 $^{210}$Po & 0.711  &  7.38  & 11.22 & $^{12}$C+$^{198}$Pt& -29.93(0)         &   -16.33 & 1.050 \\
            &        &         &      &$^{18}$O+$^{192}$Os&  -35.89(-0.78)& -16.33   & 0.982 \\
 $^{212}$Rn & 0.732   &7.83 & 8.88    &$^{18}$O+$^{194}$Pt& -34.79(-0.78)    & -9.26  & 0.970   \\
 $^{214}$Rn & 0.729   &7.54 &  9.19    &$^{16}$O+$^{198}$Pt& -29.93(-4.74)    & -4.77 & 0.996  \\
 $^{216}$Rn & 0.727   &7.25 & 9.49     & $^{18}$O+$^{198}$Pt&-29.93(-0.78)    &0.70   & 0.977   \\
 $^{213}$Fr & 0.743    &8.06 &  7.83   &$^{16}$O+$^{197}$Au&-31.16(-4.74)      &-4.01 & 0.987   \\
            &          &     &         &$^{19}$F+$^{194}$Pt& -34.79(-1.49)    &-4.01  & 0.954 \\
 $^{215}$Fr & 0.740  &7.76 &   8.13    &$^{19}$F+$^{196}$Pt& -32.67 (-1.49)   &-0.07  & 0.958  \\
 $^{217}$Fr & 0.737  &7.47  &  8.42      &$^{19}$F+$^{198}$Pt& -29.93(-1.49)    &5.00  &0.961\\

 \hline
  \hline\hline

\end{tabular}
\end{adjustbox}
\end{table*}
\begin{figure*}[t!]
\centering
            \includegraphics[width=.9\textwidth,keepaspectratio=true]{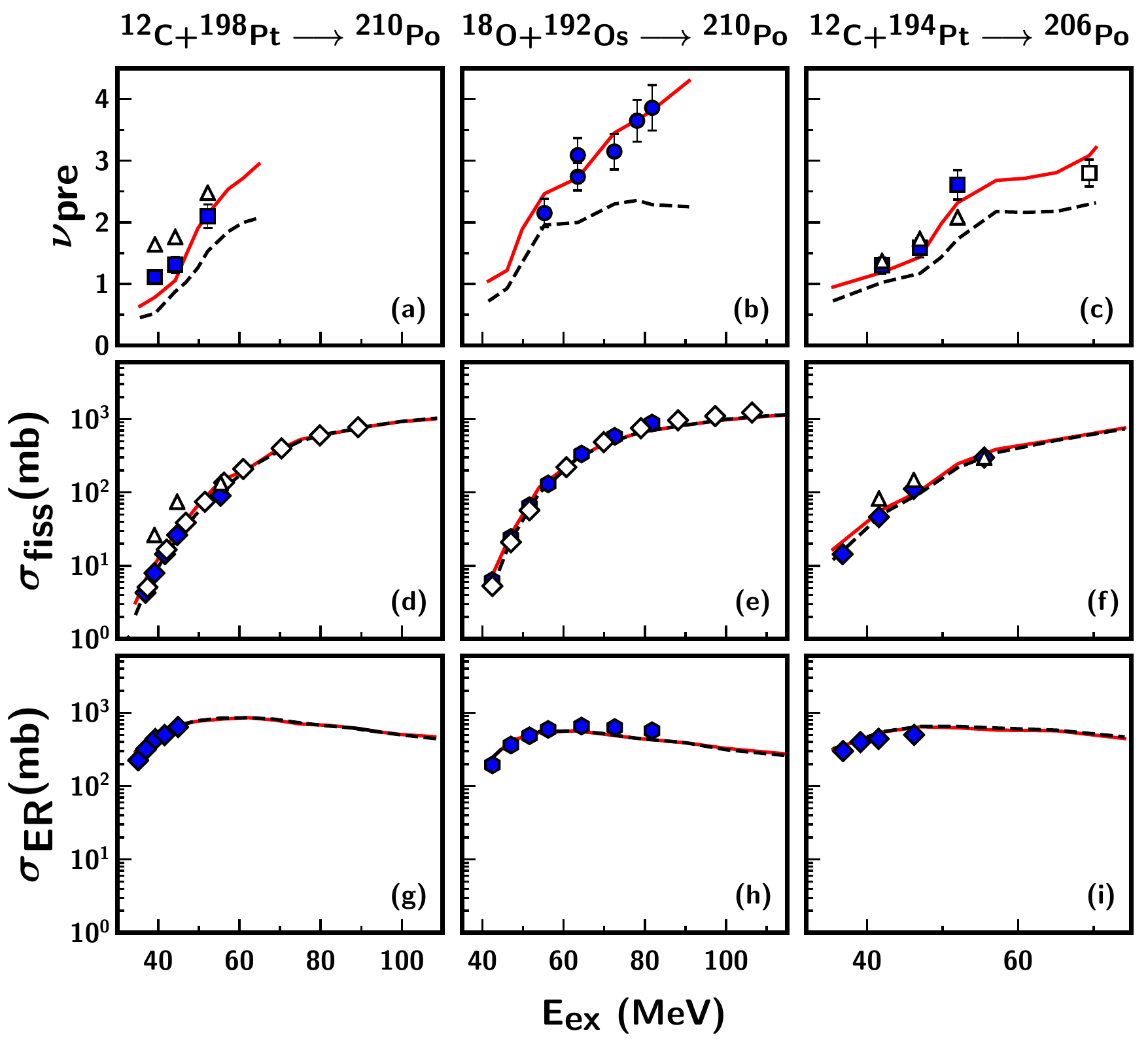}
          \caption{(Colour online) Measured and calculated pre-scission neutron multiplicities ($\textbf{$\nu_{pre}$}$), fission cross-sections ($\textbf{$\sigma_{fiss}$}$) and evaporation residue cross-sections ($\textbf{$\sigma_{ER}$}$) as a function of excitation energy for the reactions $^{12}$C$+^{198}$Pt, $^{18}$O$+^{192}$Os and $^{12}$C$+^{194}$Pt. The continuous line (red) denote calculated results with a universal frictional form factor and dashed line (black) represent statistical model calculations. The symbols in the legend represent different experimental data sets, for $\textbf{$\nu_{pre}$}$: (filled squares) Ref. \cite{Golda2013}, (filled circles) Ref. \cite{Newton1988} and (open square) Ref. \cite{Chubaryan}; for $\textbf{$\sigma_{fission}$}$ and $\textbf{$\sigma_{ER}$}$: (filled diamonds) Ref. \cite{Shrivastava1999}, (filled hexagons) Ref. \cite{Charity1986} and (open diamonds) Ref. \cite{Plicht1983}. The open triangles represent results of $\textbf{$\nu_{pre}$}$ and $\textbf{$\sigma_{fission}$}$ from 4D Langevin calculations of Ref. \cite{Schmitt2014}. }
     
 \label{fig:fig1}
\end{figure*}

Fig. \ref{fig:fig1} shows the results of dynamical calculations compared with the experimental data of $\nu_{pre}$, fission and ER cross-sections for $^{206}$Po formed via. $^{12}$C+$^{194}$Pt \cite{Golda2013,Shrivastava1999,Chubaryan} reaction and $^{210}$Po formed through two different entrance channel reactions, namely $^{12}$C+$^{198}$Pt \cite{Golda2013,Shrivastava1999,Plicht1983} and $^{18}$O+$^{192}$Os \cite{Newton1988,Charity1986,Plicht1983}, spanning a wide range of excitation energy. The excitation energies shown here are with respect to the liquid drop ground state CN mass and experimental mass of projectile and target \cite{Moller1995}.  Our calculations are restricted to excitation energies at and above 40 MeV where the present macroscopic model is valid. We emphasize that the microscopic shell corrections are not accounted for in the present calculations, as we are dealing with hot nuclei where shell effects are expected to be negligible  at high excitation energies that are populated in heavy-ion reactions. The results of calculations using only the statistical model (dashed line) are also shown in Fig. \ref{fig:fig1}. These calculations are made with the same code with Langevin dynamics turned off. The statistical model calculations under-predict the measured $\nu _{pre}$ data as shown in panels (a) to (c), even more so as excitation energy increases. The dynamical model calculations using universal reduced friction coefficient are in excellent agreement with the measured data of $\nu _{pre}$ (panels (a) to (c)), fission cross-sections $\sigma_{fiss}$ (panels (d) to (f)) and ER cross-sections $\sigma_{ER}$ (panels (g) to (i)) for the neutron shell closed nuclei $^{210}$Po as well as its isotope $^{206}$Po. The measured data of $^{210}$Po formed through two different entrance channels agree well with the theory in a broad range of excitation energies up to 80 MeV. The model calculations describe the available experimental data for $^{206,210}$Po simultaneously at these excitation energies without any microscopic corrections included in the model. These observations are at variance with the statistical model analysis of $^{12}$C+$^{194}$Pt and  $^{12}$C+$^{198}$Pt reactions that reported a significant shell correction at the saddle deformation to describe the angular anisotropy and $\nu _{pre}$ data \cite{Shrivastava1999,Golda2013}. A recent 4D Langevin dynamical study \cite{Schmitt2014} that was carried on $^{206}$Po and $^{210}$Po populated from reaction $^{12}$C+$^{198}$Pt, reported a reasonable description of the measured data for these reactions without invoking any extra shell corrections at the saddle state; shown as open triangles in panels (a), (c), (d) and (f) of Fig. \ref{fig:fig1}. A better agreement of the measured data is observed for $^{12}$C+$^{198}$Pt reaction in comparison to its 4D Langevin calculations \cite{Schmitt2014}, particularly at low excitation energies as shown in panels (a) and (d) of Fig. \ref{fig:fig1}.  The overestimation of $\nu _{pre}$ and fission cross-section of $^{210}$Po in Ref. \cite{Schmitt2014} was attributed to the remnant of ground state shells and hence, a consequence of not using a pure macroscopic potential energy surface as suggested in Ref. \cite{Mahata2017}. Nonetheless, the predictions of multi-dimensional Langevin model for $\nu _{pre}$ data of $^{206}$Po by Karpov $et$ $al$. \cite{Karpov} are also found to be in reasonable agreement with the results of the present analysis. Moreover, the measured mass distribution of fragments in the fission of $^{206,210}$Po \cite{Sen2017,Chaudhuri2015a} reaffirms the absence of any shell corrections on the potential energy surface at the saddle point.
\begin{figure*}[t!]
\centering
       \includegraphics[width=.9\textwidth,keepaspectratio=true]{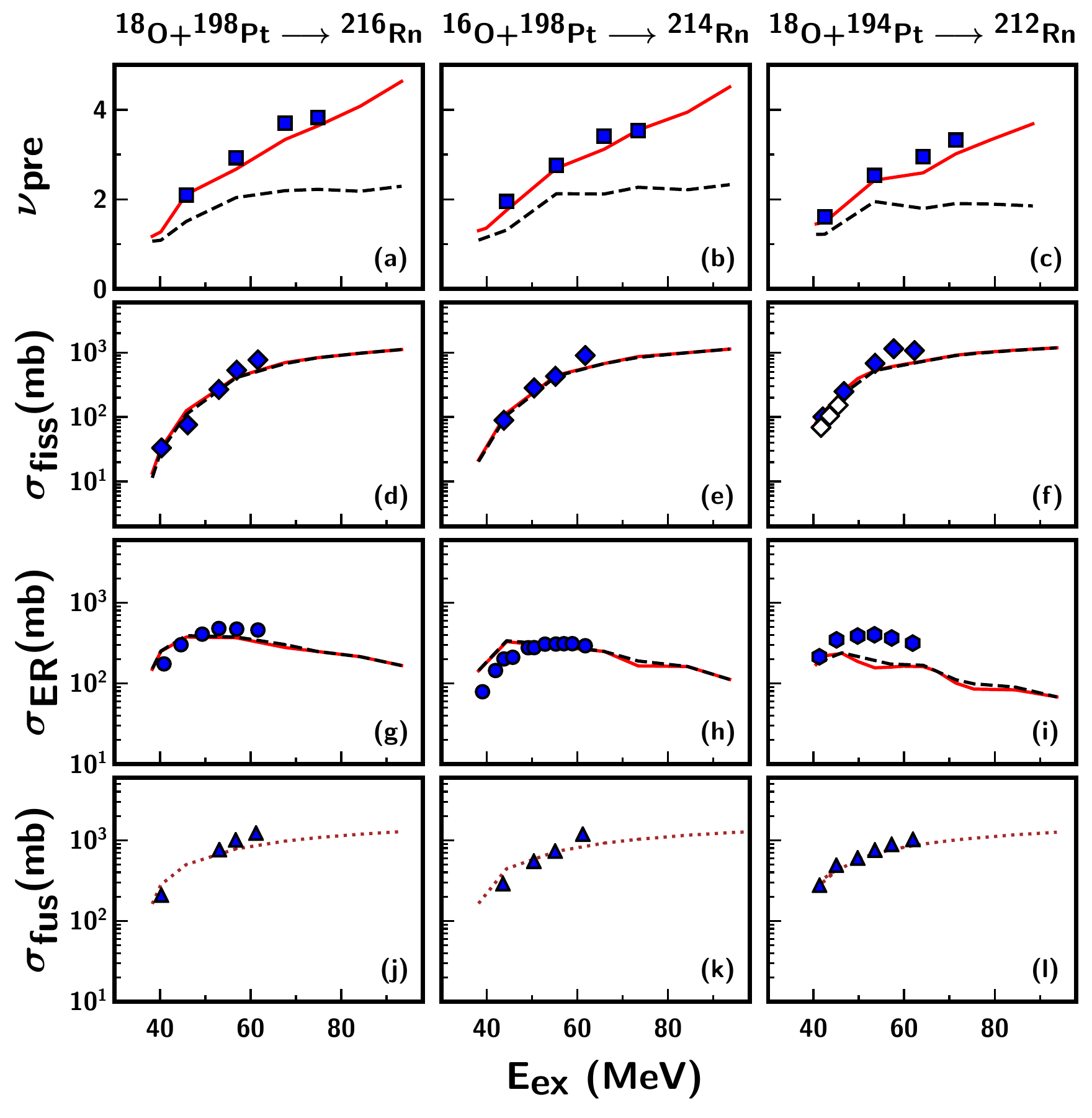}
          \caption{(Colour online) Measured and calculated pre-scission neutron multiplicities ($\textbf{$\nu_{pre}$}$), fission cross-sections ($\textbf{$\sigma_{fiss}$}$), evaporation residue cross-sections ($\textbf{$\sigma_{ER}$}$) and fusion cross-sections ($\textbf{$\sigma_{fus}$}$)  as a function of excitation energy for the reactions $^{18}$O$+^{198}$Pt, $^{16}$O$+^{198}$Pt, $^{18}$O$+^{194}$Pt. The continuous (red) and dashed (black) lines have the same meaning as in Fig. \ref{fig:fig1}. The calculations of fusion cross-section are independent of the frictional form and are represented by dotted line (brown). The symbols in the legend represent different experimental data sets, for $\textbf{$\nu_{pre}$}$: (filled squares) Ref. \cite{Sandal2013}; for $\textbf{$\sigma_{fiss}$}$: (filled diamonds) Ref. \cite{Sandal2013a} and (open diamonds) Ref. \cite{Laveen2015}; for $\textbf{$\sigma_{ER}$}$: (filled circles) Ref. \cite{Sandal2015} and (filled hexagons) Ref. \cite{Laveen2015} and for $\textbf{$\sigma_{fus}$}$: (filled triangles) Refs. \cite{Sandal2015,Laveen2015}. }    
  \label{fig:fig2}
\end{figure*}

\begin{figure*}[t!]
   \centering
 \includegraphics[width=0.9\textwidth,keepaspectratio=true]{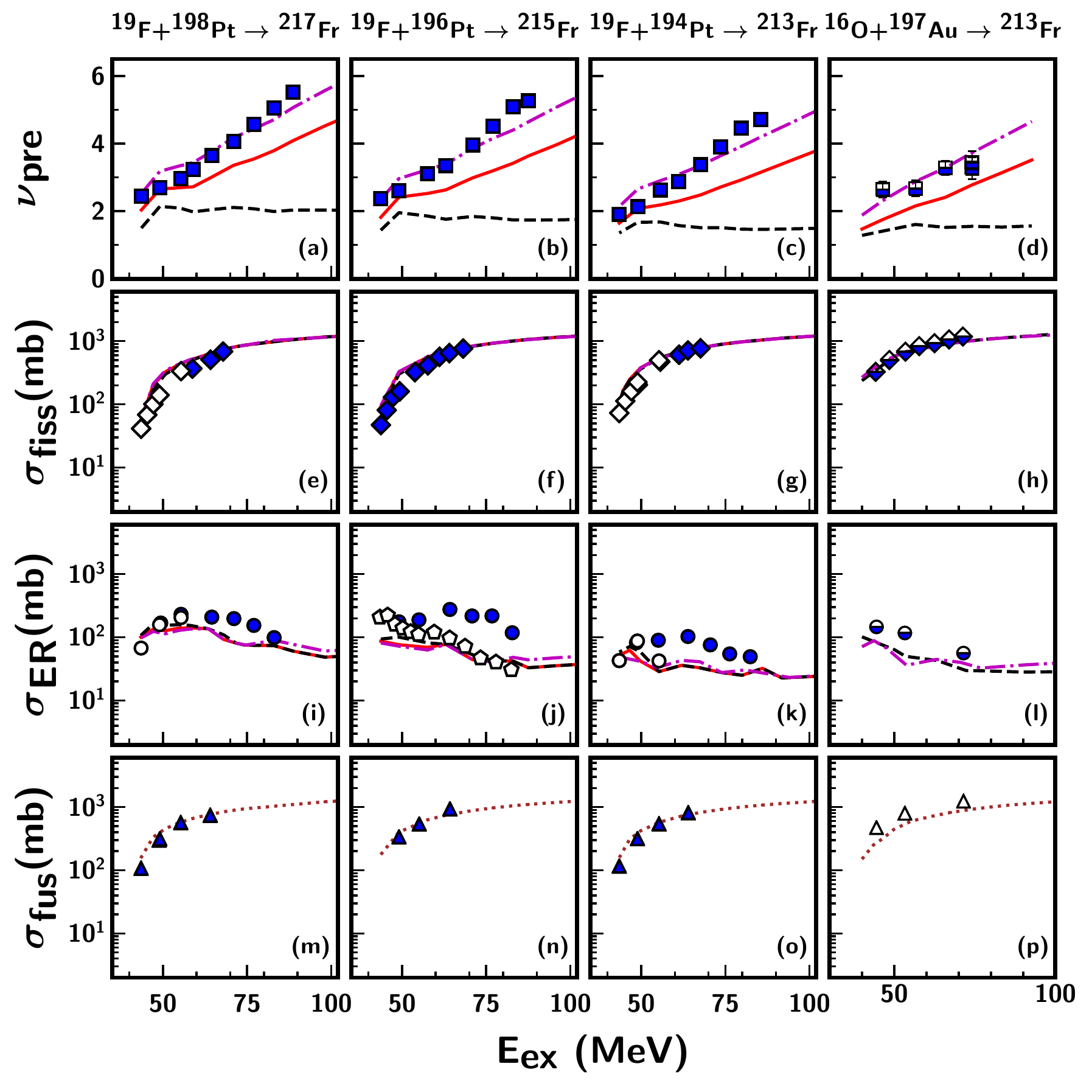}
          \caption{(Colour online)  Measured and calculated pre-scission neutron multiplicities ({$\nu_{pre}$}), fission cross-sections ($\sigma_{fiss}$), evaporation residue cross-sections ($\textbf{$\sigma_{ER}$}$) and fusion cross-sections ($\sigma_{fus}$) as a function of excitation energy for the reactions $^{19}$F$+^{198}$Pt, $^{19}$F$+^{196}$Pt, $^{19}$F$+^{194}$Pt and $^{16}$O$+^{197}$Au. The continuous (red), dashed (black) and dotted (brown) lines have the same meaning as in Figs. 1 and 2. The dash-dotted line (magenta) represent calculated results with temperature-dependent friction. The symbols in the legend represent different experimental data sets, for $\textbf{$\nu_{pre}$}$: (filled squares) Ref. \cite{Singh2012} and (partially filled squares) Ref. \cite{Newton1988} ; for $\textbf{$\sigma_{fiss}$}$: (filled diamonds) Ref. \cite{Singh2021},(partially filled diamonds) Ref. \cite{Hinde1986} and (open diamonds) Ref. \cite{Mahata2002}; for $\textbf{$\sigma_{ER}$}$: (filled circles) Ref. \cite{Singh2014}, (partially filled circles) Ref. \cite{Hinde1986} and  (open circles) Ref. \cite{Mahata2002} and for {$\sigma_{fus}$}: (filled triangles) Refs. \cite{Singh2021,Singh2014,Mahata2002} and (open triangles) Refs. \cite{Hinde1986}. The open pentagons denote $\textbf{$\sigma_{ER}$}$ for $^{215}$Fr nuclei formed via $^{18}$O$+^{197}$Au  Ref. \cite{Corradi2005}.}
\label{fig:fig3}
\end{figure*}

Figs. \ref{fig:fig2} and \ref{fig:fig3} display the comparison between experimental data and theoretical calculations  of $\nu_{pre}$, fission, ER and fusion cross-sections for N$=$126 shell closed nuclei viz. $^{212}$Rn \cite{Sandal2013,Sandal2013a,Laveen2015,Sandal2015} formed through reaction $^{18}$O+$^{194}$Pt and $^{213}$Fr formed through reactions $^{19}$F+$^{194}$Pt \cite{Singh2012,Mahata2002,Singh2014,Singh2021} and $^{16}$O+$^{197}$Au \cite{Newton1988,Hinde1986}, and their non-shell closed isotopes $^{214,216}$Rn  populated via. reactions $^{16,18}$O+$^{198}$Pt \cite{Sandal2013,Sandal2013a,Sandal2015,Laveen2015} and $^{215,217}$Fr populated via. reactions $^{19}$F+$^{196,198}$Pt \cite{Singh2012,Mahata2002,Singh2014,Singh2021}. The model calculations describe the $\nu_{pre}$ and fission excitation functions for $^{212}$Rn and its isotopes $^{214,216}$Rn quite successfully. In reactions forming $^{213,215,217}$Fr nuclei, the same parameter set is able to account for the experimental fission excitation functions but not  $\nu_{pre}$.  A recent work \cite{Singh2021} using an extended version of statistical-model employing collective enhancement of level density also reported an under-estimation of $\nu_{pre}$ data for same reactions when fitted simultaneously with fission cross-section. In the present work, the disagreement between experimental $\nu_{pre}$  and theory is prominent above $\approx$50 MeV excitation energy and it increases with rise in excitation energy. Considering that $\nu_{pre}$ of other studied nuclei are well reproduced by the model, it is unclear why the same frictional form fails, particularly for reactions forming Fr nuclei. It is to be noted that, an energy dependent dissipation was used in Ref.\cite{Sandal2013,Singh2012} to describe the $\nu_{pre}$ data for these reactions. We also attempted similar approach by employing a temperature-dependent friction (TDF) in the stochastic calculations \cite{Gontchar1997} (without changing any other parameter). This frictional form factor is deformation dependent, unlike the ones used in Refs. \cite{Sandal2013,Singh2012,Thoennessen}. The maximum of $\beta(q)$ in TDF corresponds to the ground state, that tends to decrease with increasing deformation with its minimum near the saddle configuration and is followed by an increase in the dissipation strength when approaching the scission. The dissipation coefficient assumes a higher value with increasing temperature of the CN.  It is observed that a better agreement of $\nu_{pre}$ data is achieved for reactions $^{19}$F+$^{194,196,198}$Pt and $^{16}$O+$^{197}$Au after invoking temperature dependence of the dissipation. The same frictional form, however, is found to over-predict the measured $\nu_{pre}$ data of other studied nuclei and hence is not shown here. 

Deviation in ER excitation functions are also to be noted for $^{212}$Rn and $^{213,215,217}$Fr nuclei wherein the calculated ER cross-sections underpredict the experimental data for these nuclei at high excitation energies. The case of Rn isotopes is of particular interest as the ER cross-section data for $^{214,216}$Rn \cite{Sandal2015} agrees fairly well with the model calculations at all measured energies but differ for $^{212}$Rn \cite{Laveen2015} except at the lowest energy. For $^{213,215,217}$Fr nuclei, the measured ER cross-sections of Ref. \cite{Singh2014} differ above excitation energy $\approx$ 55 MeV and the deviation is prominent for $^{213,215}$Fr. It is quite interesting to note that the ER measurement  by a different group \cite{Mahata2002} for  same reactions forming $^{213,217}$Fr at  E$_{ex}$ $\leq$ 55 MeV follows the trend of the model predictions quite successfully. Unfortunately, Ref. \cite{Mahata2002} has reported only three data points. Moreover, the ER cross-section data of $^{215}$Fr formed in reaction $^{18}$O+$^{197}$Au \cite{Corradi2005} is reproduced reasonably well with results of  $^{19}$F+$^{196}$Pt particularly, above 50 MeV excitation energy (displayed as open pentagons in panel (j) of Fig. \ref{fig:fig3}). The present dynamical calculations assume decay from an equilibrated CN and any entrance channel effects are not included. It takes account of only the different angular momenta that are populated in different entrance channels. Taking into consideration the insignificant difference in  angular momenta between two entrance channels forming $^{215}$Fr, the observed deviation in ER cross-section for $^{19}$F-induced reaction is quite unexpected. These observations further necessitated the need to confront the deviations  in describing ER cross-sections by comparing the measured fusion cross-sections for Rn and Fr nuclei with the model. It is revealed that the calculated fusion cross-sections are in good agreement with the measured fusion data, augmenting the validity of the present calculations. Furthermore, the under-prediction of ER cross-sections indicates the need for a strong dissipation in the pre-saddle region \cite{Back1999}. However, 3D Langevin dynamical calculations \cite{Nadtochy2002} reported  a reduction in the wall friction coefficient to reproduce the mass and kinetic energy distribution of fission fragments, and their influence on $\nu_{pre}$ for $^{215}$Fr nucleus. The strength of the reduction coefficient, k$_s$ $=$ 0.25 $-$ 0.5 indicates a weak dissipation in the initial stages of the fissioning nucleus. The experimental analysis of fission fragment nuclear-charge distributions and fission cross-sections of Fr, Rn isotopes and their neighbouring nuclei also reported a pre-saddle dissipation strength  of magnitude (4.5 $\pm$ 0.5) $\times$ 10$^{21}$ s$^{-1}$ \cite{Schmitt2007} and 2 $\times$ 10$^{21}$ s$^{-1}$ \cite{Jurado2004}, respectively. The more recent microscopic study of energy dependent dissipation using time-dependent Hartree-Fock + BCS method \cite{Qiang2021} also observed a strength of deformation dependent friction coefficient, ranging from 1 to 6 $\times$ 10$^{21}$ s$^{-1}$ in heavy nuclei. The strength of these frictional parameterizations are quite in agreement with the dissipation form factor employed in the present calculations. These observations affirm a weak dissipation in the pre-saddle region; so, the observed enhancement of ER cross-sections in Fr nuclei populated via. $^{19}$F-induced reactions is not well-understood from the perspective of dissipation strength alone. In fact, a satisfactory description of the excitation functions including ER cross-sections for reactions $^{12}$C+$^{194}$Pt, $^{12}$C+$^{198}$Pt, $^{18}$O+$^{192}$Os and $^{16,18}$O+$^{198}$Pt and survival probabilities for a range of fissilities \cite{Frobrich1998} is observed within the framework of this 1D Langevin dynamics with a universal friction parameter. However, it is also important to bear in mind the possible bias coming from experimental uncertainty. It is striking that the observed deviations are pronounced in ER cross-section data where measurements are reported to have large uncertainty in ER separator transmission efficiency \cite{Laveen2015,Singh2014}. It would be highly desirable to have additional ER measurements to rule out any possible experimental bias in the interpretation of ER data.

\begin{figure}
\centering
           \includegraphics[width=.45\textwidth,keepaspectratio=true]{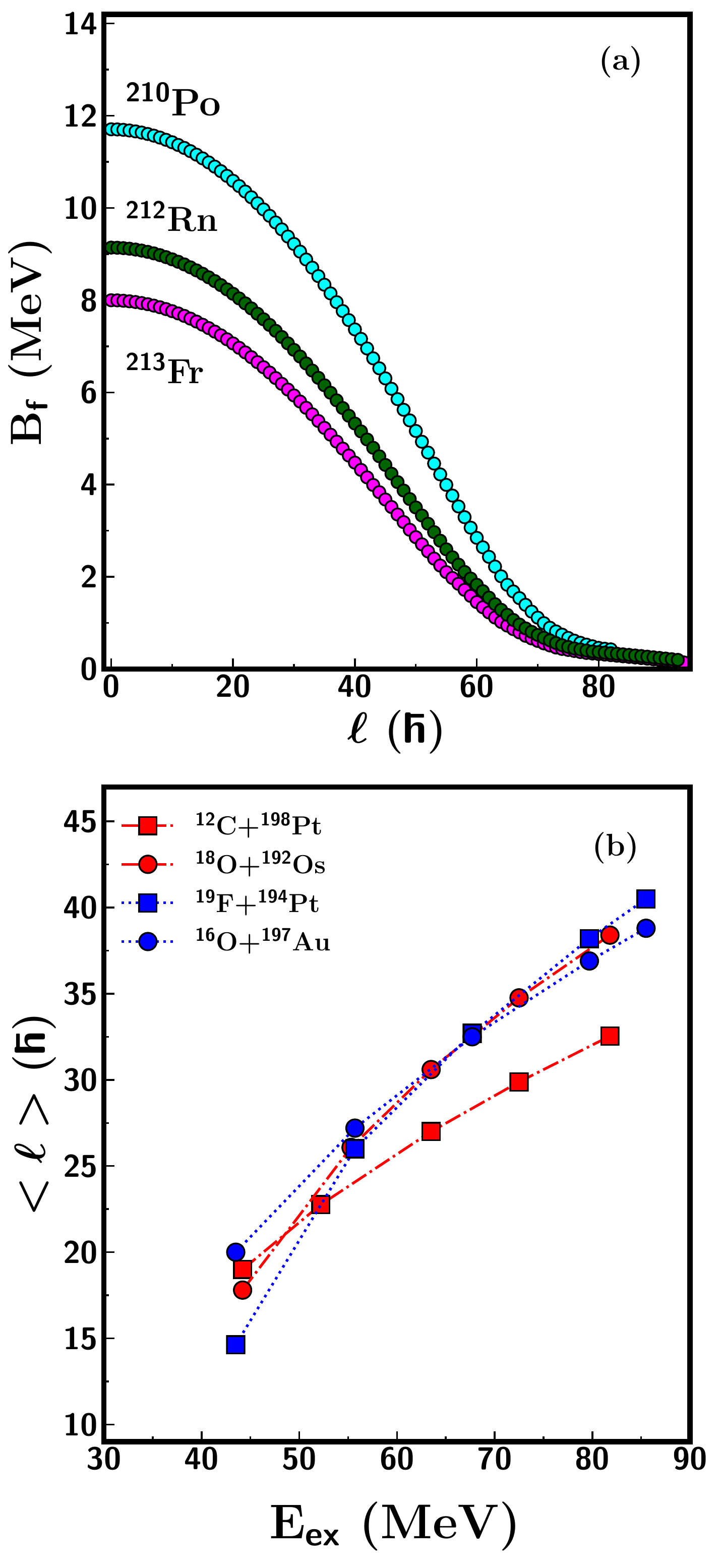}
    \caption{(Colour online) (a) The angular momentum '$l$' dependent fission barrier height $B_f(l)$  for three CN $^{210}$Po,$^{212}$Rn and $^{213}$Fr and (b) Variation of  mean angular momentum $<l>$ with compound nucleus excitation energy for $^{210}$Po,and $^{213}$Fr populated by different entrance channels. }
     \label{fig:fig4}
\end{figure}

 It must be noted that, the entrance channel dynamics of the fusion stage might also play a role influencing neutron emission at the formation stage \cite{Saxena1994}. It is known that interplay of  CN excitation energy, angular momentum and fission barrier play crucial role in fission process \cite{Schmitt2018}. Present study do not take into account any entrance channel dynamics influencing the fusion stage. The model only considers the entrance channel dependent '$l$' distribution  calculated within the surface friction model \cite{Marten1992}. In Fig. \ref{fig:fig4} we show the calculated fission barrier height $B_f(l)$ for three compound systems and  mean angular momentum $<l>$ calculated from '$l$' distribution for different entrance channels forming same CN. The variation of $B_f$ is plotted as a function of '$l$' in Fig. \ref{fig:fig4}(a) and variation of  $<l>$  of the compound systems is plotted as a function of $E_{ex}$ in Fig \ref{fig:fig4}(b). From Fig. \ref{fig:fig4}, it is clear that,  the  difference in  angular momenta between two entrance channels forming same CN at similar $E_{ex}$ is not very significant to cause any '$l$' induced effects on measured fission observable.   This is evident in the $\nu_{pre}$ data for $^{210}$Po formed in reactions $^{12}$C+$^{198}$Pt and $^{18}$O+$^{192}$Os  which are well described in the present work (see Fig. \ref{fig:fig1}) without invoking any entrance channel effects in the model.  Recent studies investigating  entrance channel dynamics \cite{Shareef2016,Schmitt2018} reported disagreement between experimental $\nu_{pre}$ and predictions  of entrance channel model for  $^{213}$Fr nuclei formed via. $^{16}$O+$^{197}$Au and $^{19}$F+$^{194}$Pt reactions.  These studies were, however, not extended to other isotopes of Fr, namely $^{215,217}$Fr that also show similar discrepancy as reported in the present study.

 \begin{figure}
\centering
           \includegraphics[width=.45\textwidth,keepaspectratio=true]{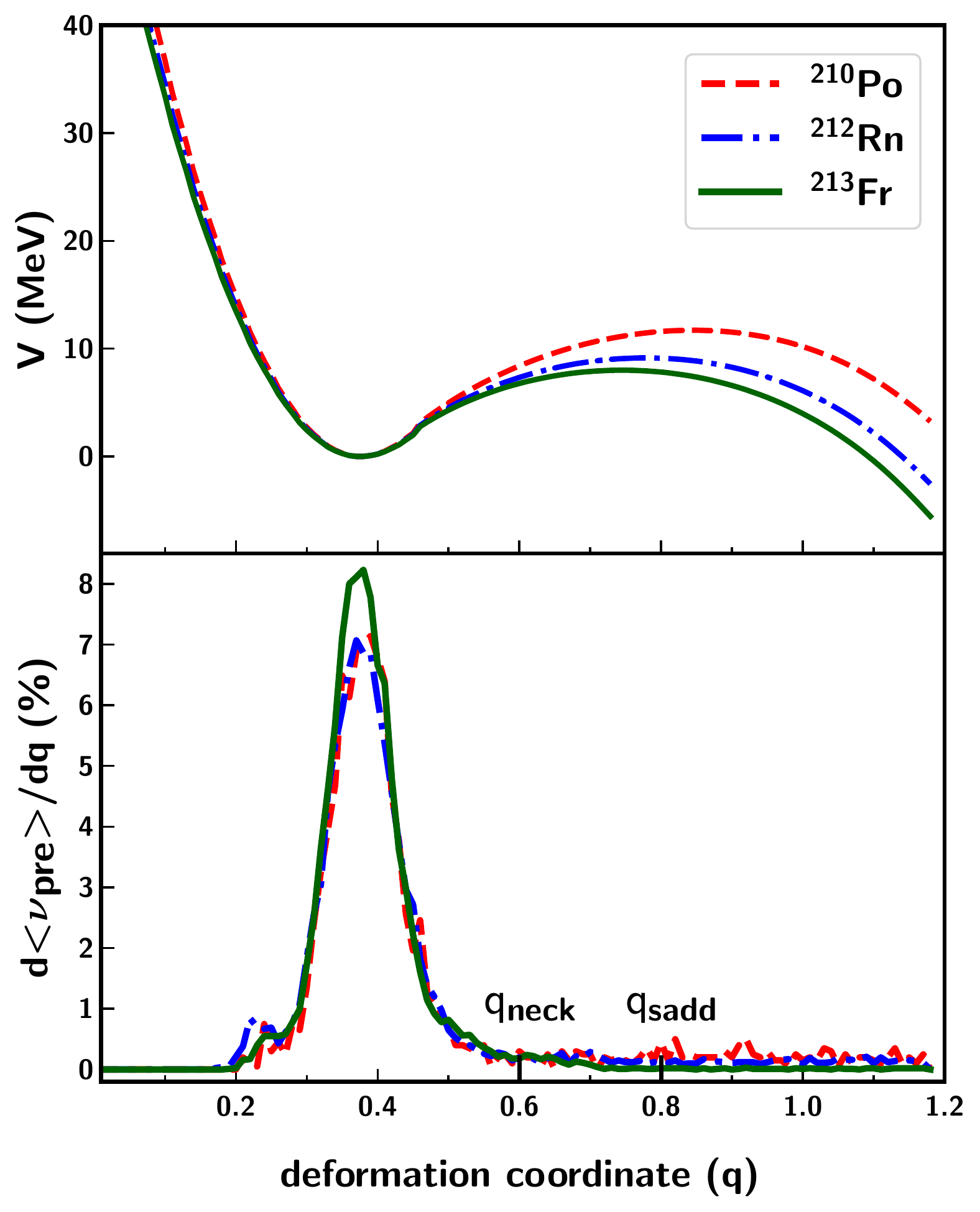}
    \caption{(Colour online) Potential energy distribution as a function of nuclear deformation coordinate (q) for three fissioning nuclei $^{210}$Po, $^{212}$Rn and $^{213}$Fr (top panel)  and distribution of percentage yield of evaporated pre-scission neutrons as a function of (q) for three CN at 50 MeV excitation energy (bottom panel). The deformation coordinate (q) assumes a value of 0.6 (q$_{neck}$) when the neck of the fissioning nucleus starts to develop and q=0.8 (q$_{sadd}$) at the saddle state configuration.}
     \label{fig:fig5}
\end{figure}

The current 1D Langevin analysis provides a simultaneous description of the   experimental data for neutron magic nuclei $^{210}$Po  without invoking any saddle shell corrections or a  nuclear dissipation  strength dependent on system/observable under study. In order to understand qualitatively that consideration of saddle shell corrections are not required to explain $\nu_{pre}$ data, we consider the nature of neutron emission during the fission process. It is to be noted that these neutrons are emitted from dynamical trajectories that originated from compact configuration till scission point is reached. The prompt and beta-delayed neutron emissions from fission fragments are not taken into consideration. As recent publications have advocated for the inclusion of shell corrections on the saddle configuration to describe the angular anisotropy and $\nu_{pre}$ data at moderate excitation energies \cite{Mahata2006,Shrivastava1999,Golda2013}, we have attempted to find the distribution of pre-scission neutrons as it evolves from ground state to scission point. The model calculated potential energy V(q) and distribution of percentage yield of pre-scission neutrons  are plotted as a function of the deformation coordinate (q) for these nuclei at 50 MeV excitation energy and shown in Fig. \ref{fig:fig5}. It is evident that more than 90$\%$ of the neutron emission occurs at an early stage of fission before the saddle deformation (q $\approx$ 0.8) \cite{Gontchar1993} is reached.  The mean of the distribution corresponds to $\nu_{pre}$  emission close to the ground state configuration. In-fact, a multi-dimensional Langevin study of $^{215}$Fr by Nadtochy $et$ $al$. \cite{Nadtochy2002} have also pointed out that an appreciable part of pre-scission neutrons are emitted at an early stage of fission before saddle is reached. As most of the neutrons are emitted close to the ground state configuration, it is unlikely to be influenced by any shell corrections applied at the saddle.

\begin{figure}
   \centering
      \includegraphics[width=.45\textwidth,keepaspectratio=true]{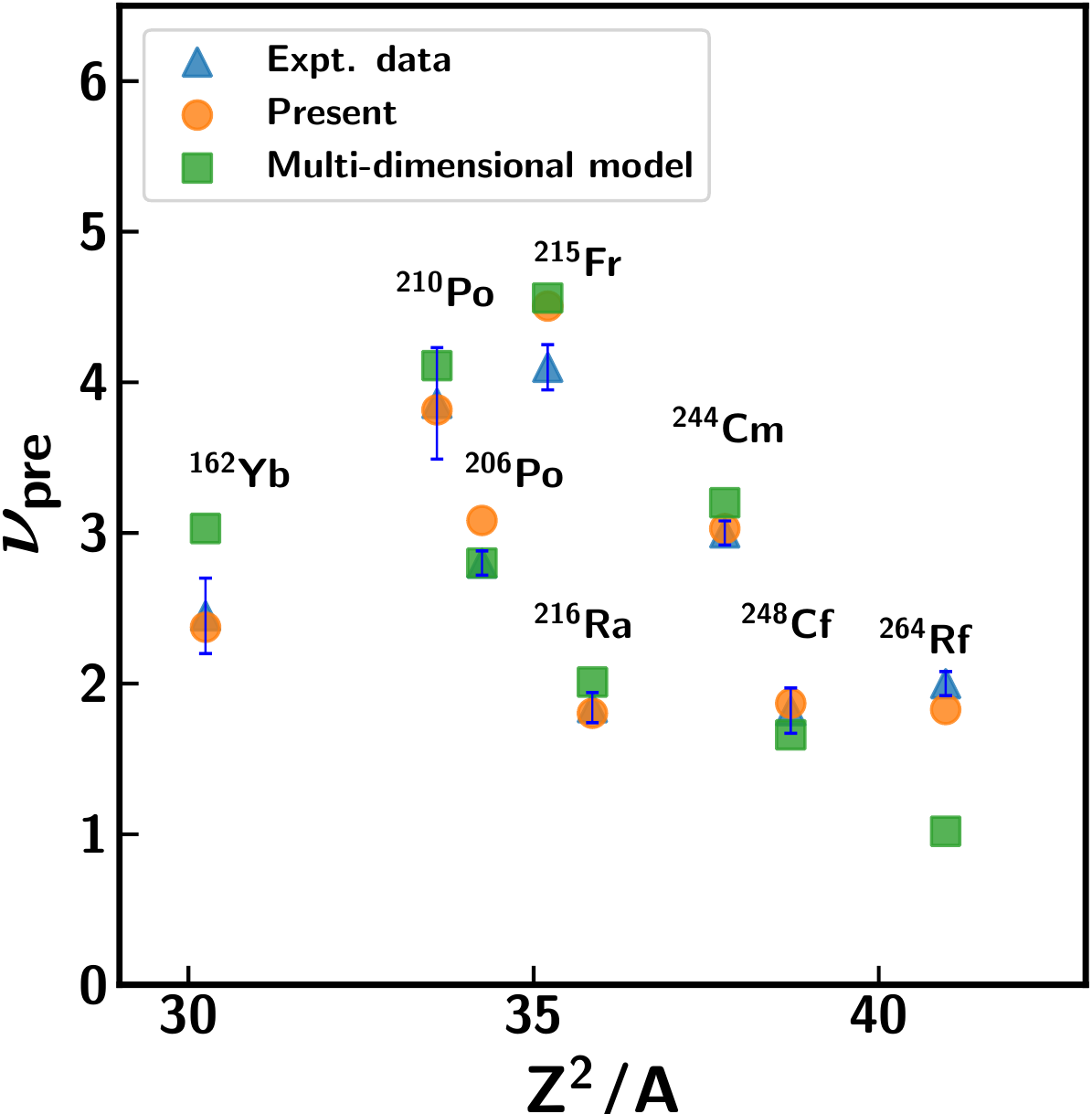}
          \caption{ (Colour online) Comparison of measured pre-scission neutron multiplicities ($\nu_{pre}$) with the results of the 1D model (present work) and multi-dimensional models. The filled triangles (blue) denote experimental data \cite{Hinde1992,Hinde1986,Chubaryan,Singh2008,Itkis1990,Saxena1994,Ferguson}, the present dynamical model calculations are represented by filled circles (orange) and the filled squares (green) denote the results of multi-dimensional dynamical calculations \cite{Nadtochy2014,Eslamizadeh2011,Karpov,Schmitt2018}.}
        
\label{Fig:Fig6}
\end{figure}
 
Though the present code uses classical 1D approach to describe fission observables, the main objective of this work is to have a simultaneous description of experimental data without any parameter adjustment thus, removing some of the reported ambiguities.  A comparison between $\nu_{pre}$ calculated with 1D model and recent macroscopic multi-dimensional models is displayed in Fig. \ref{Fig:Fig6}. It can be seen that the $\nu_{pre}$ values predicted by different models are very similar and also reproduce the measurements quite well for reactions spanning a wide range of fissility parameter $Z^{2}/A$. Additionally, the multi-dimensional  calculations \cite{Nadtochy2012,Nadtochy2014,Chaudhuri2002} also use the formalisms adopted from Refs. \cite{Gontchar1997,Frobrich1998} such as the parameterization of surface friction model and weakest coordinate dependence of the level-density parameter as employed in the present work. Hence, the qualitative nature of the observed features presented here is not expected to be different with multi-dimensional approach.  As the present framework is found to provide realistic values close to measured data, we believe that the 1D approach  still can be a potential tool to study a wider systematics which can be accomplished within minimum computational resources.

It must be remarked here that, even though present analysis provides a reasonable reproduction of the experimental data without invoking any shell corrections at high  excitation energies, it shall not be concluded from this work that shell effects are not relevant in the analysis. As present investigation consider only the first chance fission at $E_{ex}$ $\sim$ 40 MeV and above where shell effects are expected to be washed out, no indication for the need of including shell corrections was found. However,  for the case when the CN is populated at low excitation energies or reaches low excitation energy due to neutron emission as a consequence of competition between neutron evaporation and fission (multi-chance fission), the microscopic effects are required to be taken into consideration. Recent  microscopic study of dissipation within  Hartree-Fock + BCS framework \cite{Qiang2021} have shown a strong dependence of dissipation on  deformation and initial excitation energies of the hot nuclei. Possible influence of microscopic temperature dependence of fission barrier height and its curvature were also  emphasized in some recent studies of fully microscopic description of fission process  \cite{Zhu2016,Qiao2022}. A microscopic framework based on the finite-temperature Skyrme-HartreeFock+BCS approach \cite{Goodman1981} was adopted to demonstrate the essential role of energy dependent fission barriers by studying the experimental fission probability of $^{210}$Po. It would be quite interesting to extend the investigation of Fr nuclei within such a microscopic framework.  

\section{Summary and Conclusion}
In the present work we report a systematic study on the fission dynamics of N=126 shell closed nuclei in mass region 200 with a simultaneous description of three fission observables. The present work highlights the limited reliability of the conclusions drawn from the recent statistical model analysis of shell closed nuclei, namely $^{210}$Po, $^{212}$Rn and $^{213}$Fr at excitation energies 40 MeV and above, that advocated for extra shell effects at saddle configuration even after their inclusion in the level density formulation. Earlier analyses of $\nu_{pre}$  and ER cross-sections were based on different assumptions and case dependent parameter adjustments, without reaching a definite conclusion. On the basis of present analysis we conclude that, without many of those assumptions and parameter adjustments, a well established combined dynamical and statistical model can simultaneously reproduce the available data of $\nu_{pre}$, fission and evaporation residue  excitation functions (also fusion cross-sections in certain cases) for neutron shell closed nuclei, viz. $^{210}$Po, $^{212}$Rn and  their non-shell closed isotopes $^{206}$Po and $^{214,216}$Rn without the need of including any extra shell effects.  There appears to be no discernible influence of N$=$126 neutron shell structure on these measured fission observables in the medium excitation energy range. The present work also points to a relatively smaller role of entrance channel effects in the studied systems.

However, we find a significant mismatch between measured $\nu_{pre}$ data and its model predictions for Fr nuclei formed in reactions $^{19}$F+$^{194,196,198}$Pt and $^{16}$O+$^{197}$Au, despite a reasonable description of fission and fusion cross-sections. The $\nu_{pre}$ data in Fr nuclei could only be reproduced after invoking a temperature dependent frictional form. The difficulty in completely reproducing some specific measurements of Fr nuclei still remains not well-understood and additional measurements are desired.  Although the present work is limited to the study of three fission observables, it would also be interesting to extend the systematic study using recent microscopic theory within  Hartree-Fock + BCS framework.

\section{Acknowledgments}      
We are thankful to K. S. Golda and N. Saneesh for fruitful discussions. One of the authors (D.A.) acknowledges the financial support in the form of  research fellowship received from the University Grants Commission (UGC).

\section*{References}  

\end{document}